\documentclass[12pt]{article}
\usepackage{YDMutaf_pic04}
\usepackage{hyperref}
\usepackage{url}
\usepackage{graphicx}

\newcommand{\met}   {\mbox{${\hbox{$E$\kern-0.63em\lower-.18ex\hbox{/}}}_{T}$} }
\newcommand{\metvec}{\mbox{${\hbox{$\vec{E}$\kern-0.63em\lower-.18ex\hbox{/}}}_{T}\,$} }
\newcommand{\metx}  {\mbox{${\hbox{$E$\kern-0.63em\lower-.18ex\hbox{/}}}_{x}\,$} }
\newcommand{\mety}  {\mbox{${\hbox{$E$\kern-0.63em\lower-.18ex\hbox{/}}}_{y}\,$} }

\newcommand{\gev}   {~GeV~}
\newcommand{\dO}    {D\O~}
\newcommand{\Z}     {$Z$~}
\newcommand{\Zb}     {$Z+b$~}

\newcommand{\lumi}    [1] {#1 $pb^{-1}$}

\begin{document}

\title{\bf MEASUREMENT OF THE RATIO OF INCLUSIVE CROSS SECTIONS 
	$\sigma(p\bar{p} \to Z+b)$/ $\sigma(p\bar{p} \to Z+j)$ at \dO RunII }
\author{
Y.~D.~Mutaf \\ 
{\em Physics \& Astronomy Department, SUNY at Stony Brook, NY 11794-3800} \\
\\
(on behalf of \dO collaboration) \\
}
\maketitle

\baselineskip=14.5pt
\begin{abstract}
We study the $b$-jet production in association with \Z boson at \dO using Run II data collected between August 2002 and September 2003 corresponding to about \lumi{180} of recorded luminosity. We also present a measurement of the ratio of cross sections for $Z+b$ to $Z+j$ inclusive jet productions.

\end{abstract}

\baselineskip=17pt

\section{Introduction}
The production of $b$ quark(s) associated with the EW bosons constitutes the essential signature for several processes currently studied at \dO, like Higgs boson production. Understanding of all the processes giving rise to these signatures have paramount importance for the completeness of the Higgs searches but currently very limited due to the low production cross sections. Observation of events with $b$-jets along with $W/Z$ bosons is therefore very important for our understanding of the processes giving rise to the event signature for Higgs production. 

In this paper, we outline a study of the $b$-jet production associated with \Z boson and in particular present a measurement of the ratio of inclusive cross sections $\sigma(p\bar{p} \to Z+b)$/ $\sigma(p\bar{p} \to Z+j)$ in both the dimuon and dielectron channels of \Z .

\section{Data Sample and Event Selection}
The analysis is based on data collected with \dO Run II detector \cite{run2det} corresponding to a total recorded luminosity of about \lumi{180} in both dimuon and dielectron channels. Although the reconstruction of \Z is different, the jet selection and b-tagging performed are same in these two channels.

\par For selecting $Z \to ll$ events, we require that at least two isolated leptons with $p_T<$ 15\gev and $|\eta|<$ 2.0 must be present in the event with the dilepton invariant mass in between [80.0, 100.0]\gev for dielectrons and [65.0, 115.0]\gev for dimuons. The electrons are reconstructed via energy depositions detected in the EM layers of the calorimeter and one of the electrons is also required to have a matching central track. On the other hand, muons are detected using the hits in the muon detector with matching central tracks. 

\par
\begin{figure}[htbp]
\centerline{\hbox{ \hspace{0.2cm}
	\includegraphics[width=6.5cm]{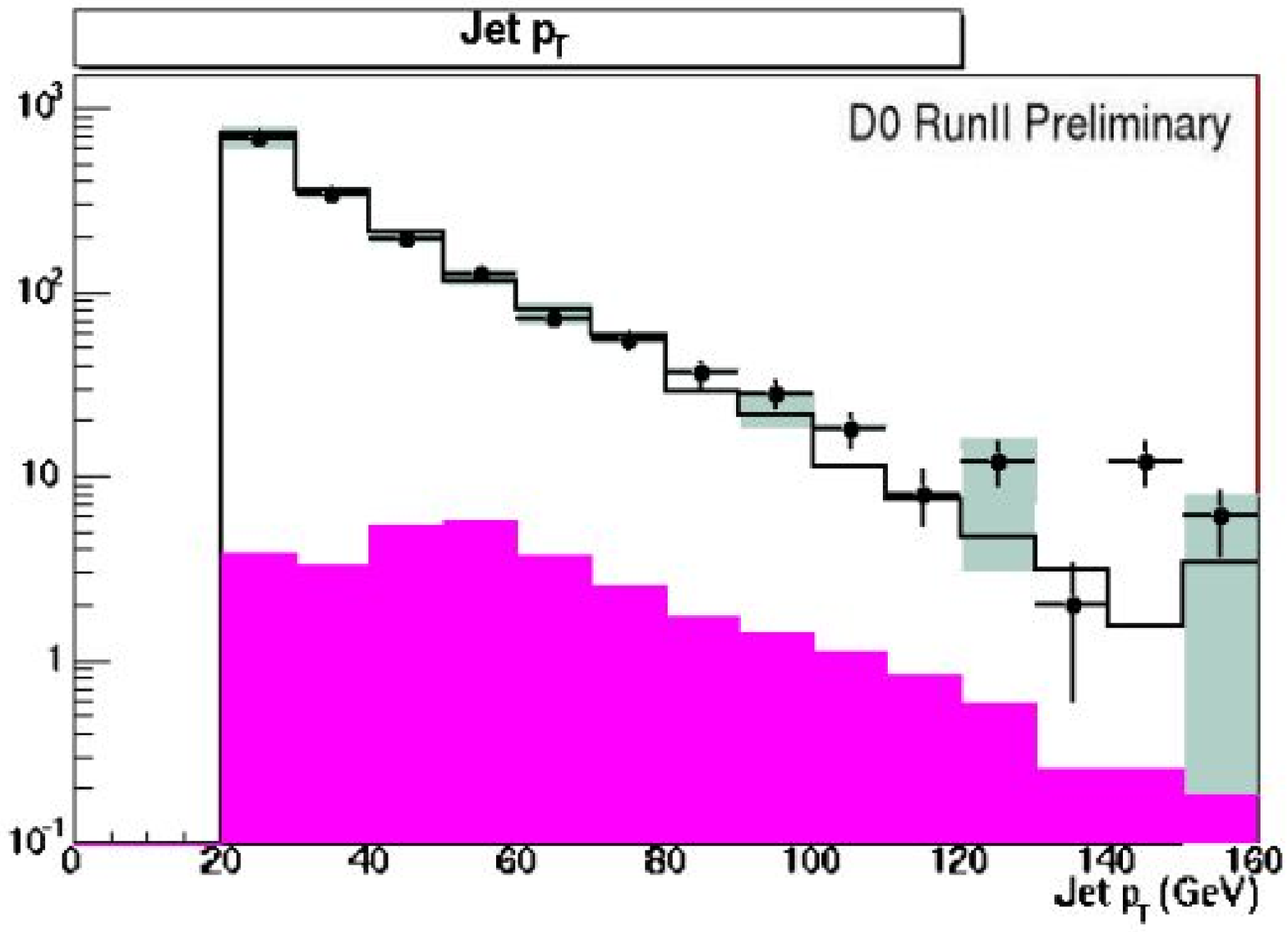}
	\hspace{0.3cm}
	\includegraphics[width=6.5cm]{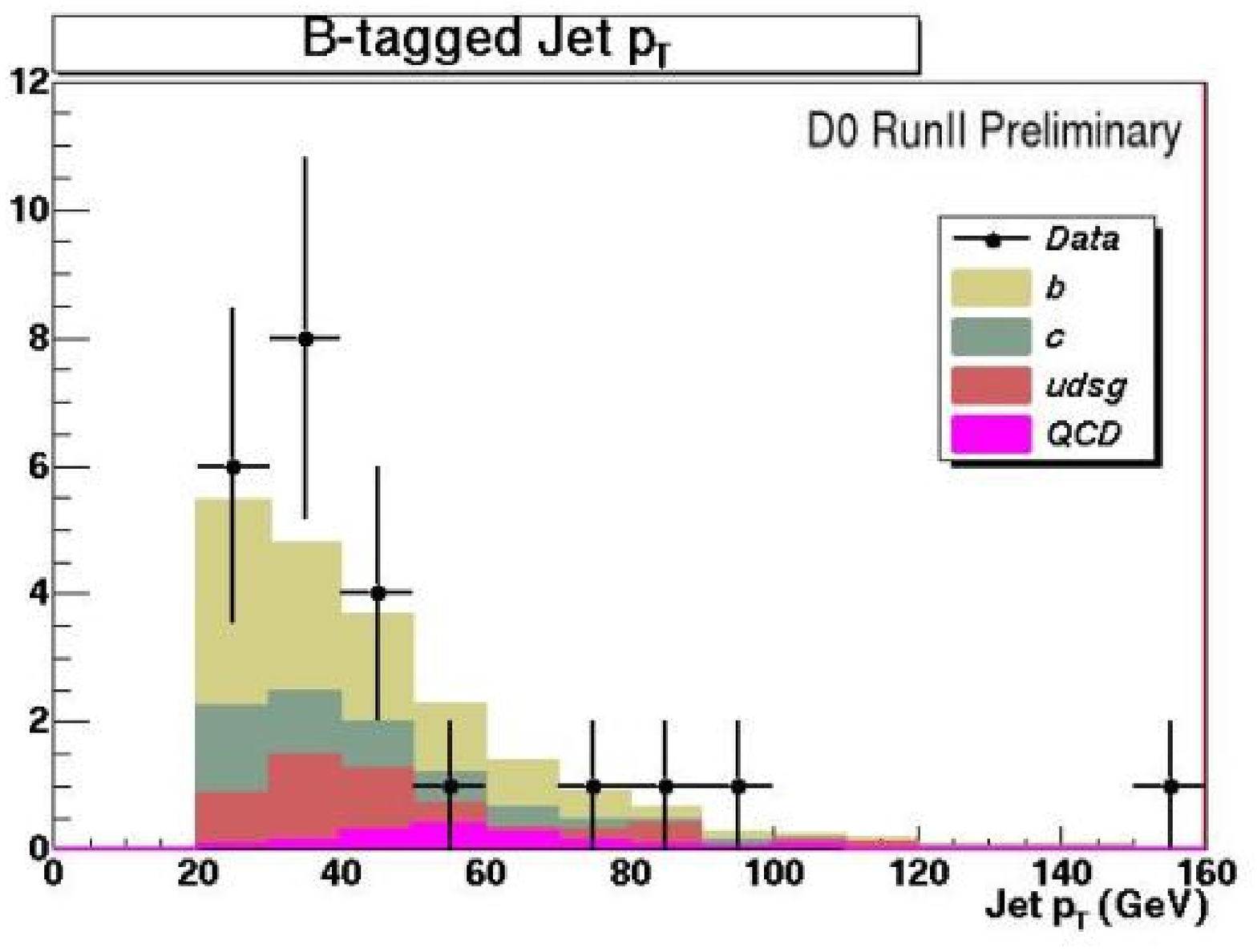}
    		}
 	 }
\caption{\it $E_T$ distribution of all taggable jets (left) and b-tagged jets (right) in selected dimuon events. Jets from data are shown with error bars and only on left with filled error boxes corresponding to systematic uncertainty due to energy scale corrections. The overlayed histograms are the normalized expectations from Z+j Alpgen (left) and Z+b Pythia MC (right). }
\label{fig:jetpt}
\end{figure}

 Furthermore, calorimeter jets in this analysis are reconstructed with Run II cone algorithm of $\Delta$R = 0.5. The reconstructed jets are required to have transverse energy $E_T<$ 20\gev and have pseudorapidity $|\eta_j| < 2.5$. Before applying b-tagging, we initally require that the selected calorimeter jets must be matched to track jets\footnote{Track jets are formed with a cone clustering algorithm of size $\Delta$R=0.5.} within $\Delta$R $<$ 0.5 (i.e. \emph{taggable} jets). Finally, we use the reconstructed secondary vertices (SV) to identify the $b$-jets among all the taggable jets\footnote{The taggable jets are identified as b-tagged if the momentum vector of SV is within $\Delta$R $<$ 0.5 of the jet axis.}. The transverse momentum distributions of taggable jets and b-tagged jets from dimuon sample are shown in Figure \ref{fig:jetpt}.

\section{Analysis}
Number of events left after \Z selection and jet requirements are shown in Table \ref{tab:number_events} for dielectron and dimuon channels. After subtracting the contributions from multijet background and Drell-Yan continuum, we can write equations relating the number of events before; $N_{before} = t_b N_b + t_c N_c + t_l N_l$, and after b-tagging; $N_{after} = \epsilon_b t_b N_b + \epsilon_c t_c N_c + \epsilon_l t_l N_l$, with different quark flavor contributions in the selected events. The symbols $t$ and $\epsilon$ in these equations refer to the corresponding b-quark, c-quark and light-parton (udsg) efficiencies for taggability and b-tagging requirements respectively. There are three unknowns in these equations, i.e. $N_b$, $N_c$ and $N_l$ and the two equations above are mathematically insufficient to solve for all of them. Therefore, we introduce an additional equation by fixing $N_b/N_c$ ratio to most recent QCD calculations \cite{theory}, i.e. $N_c = 1.69 N_b$.

\begin{table}[htbp]
\centering
\caption{ \it Number of events left after successive selection of events in the \Z mass region for dielectron and dimuon channels.}
\vskip 0.1 in
\begin{tabular}{|l|c|c|} \hline
			& $Z \to ee$ 	& $Z \to \mu\mu$	\\ \hline \hline
 \Z inclusive		& 15613		& 11543			\\
 $Z+\ge$ 1 jet		& 2219		& 1754			\\
 $Z+\ge$ 1 taggable jet	& 1658		& 1406			\\
 $Z+\ge$ 1 b-tagged jet	& 27		& 22			\\ \hline
\end{tabular}
\label{tab:number_events}
\end{table}

\section{Results}
We observe \Zb production at \dO and measure the following preliminary ratio of inclusive cross sections. 
\begin{equation}
	\frac{\sigma(p\bar{p}\to Z+b)}{\sigma(p\bar{p}\to Z+j)} = 0.024 \pm 0.005(stat)^{+0.005}_{-0.004} (syst) 
\nonumber
\end{equation}

\par The theoretical expectation for the same ratio is about 0.020 (\cite{theory}) and is consistent with our measurement.


\begin{thebibliography}{99}
  \bibitem{run2det} V. Abazov, et al., in preparation for submission to Nucl. Instrum. Methods A; T. LeCompte and H.T. Diehl, Ann. Rev. Nucl. Part. Sci. {\bf 50}, 71 (2000).
\bibitem{theory} J.M.~Campbell, R.K.~Ellis, F.~Maltoni and S.~Willenbrock, Phys. Rev. D69 (2004), 074021.

\end{thebibliography}
\end{document}